fahadabdullaah@gmail.com

# Impacts of different cumulus physics over south Asia region with case study tropical cyclone Viyaru

[1]Abdullah Al Fahad, [2]Tanvir Ahmed

[1,2] Department of Physics, Shahjalal university of science and technology, Sylhet, Bangladesh

## ABSTRACT

Tropical Cyclone Viyaru, formerly known as Cyclonic Storm Mahasen was a rapidly intensifying, category 01B storm that made landfall in Chittagong, Bangladesh on the 16th of May, 2013. In this study, the sensitivity of numerical simulations of tropical cyclone to cumulus physics parameterizations is carried out with a view to determine the best cumulus physics option for prediction of the cyclone's track, timing, and central pressure evolution in the Bay of Bengal. For this purpose, the tropical cyclone Viyaru has been simulated by WRF-ARW v3.4.1 in a nested domain with NCEP Global Final Analysis (FNL) data as initial and boundary conditions. The model domain consists of one parent domain and one nested domain. The resolution of the parent domain is 36 km while the nested domain has a resolution of 12 km. Five numerical simulations have been done with the same microphysics scheme (WSM3), planetary boundary layer scheme (YSU), NOAH land surface scheme but different Cumulus Parameterization scheme. Four cumulus Parameterization schemes are KF, BMJ, GF and Tiedtke and one simulation was done without any cumulus physics scheme. The results of model simulations are compared with corresponding analysis or observation data. For best result data provided by Joint Typhoon Warning Centre (JTWC) and NASA tropical cyclone centre was used as observed for comparison. After the study it was found that tracks, intensity, wind speed, precipitation, and central pressure of the cyclone have sensitive result with different cumulus physics schemes.

## 1. INTRODUCTION

To demonstrate impacts of the different parameterization scheme over cyclone Viyaru, weather research and forecasting (WRF) was used. WRF is a numerical weather prediction and atmospheric simulation system designed for research and operational application. Different parameterization schemes like microphysics, Cumulus parameterizations, surface physics, planetary boundary layer physics, and atmospheric radiation physics schemes are integrated with advance research WRF model. Cumulus physics schemes are adjustment and mass flux schemes which used for mesoscale modeling. Cumulus schemes used to determine when to trigger a convective column and how fast to make convective act for the WRF model. Selected convectively unstable schemes column work on that individually. Mass-flux type schemes transport surface air to top of cloud and include subsidence. With time Subsidence around cloud warms and dries troposphere by removing instability. Cumulus (CU) physics provide atmospheric heat and moisture/cloud tendency profiles, Surface sub-grid-scale (convective) rainfall.

Different studies showed cumulus microphysical processes have significant impact on numerical simulations of clouds and precipitation (Brown and Swann, 1997; McCumber et al., 1991; Meyers et al., 1992).

Corresponding author
Email address: fahadabdullaah@gmail.com
(Abdullah Al Fahad) (*Article is submitted to springer for evaluation*). Details version will be available soon.

In depth treatment of cumulus physics schemes tends to produce more accurate result and improvement on simulation significantly even in numerical models with horizontal resolution as large as 20 km (Rosenthal, 1978; Wang, 2002; Yamasaki, 1977; Zhang et al., 1988). In a study (McCumber et al., 1991) it was found that, in a numerical cloud-resolving model use of three ice classes (cloud ice, snow, graupel/hail) produced better results than did using two ice classes (cloud ice and snow) which in turn was better than no ice (warm rain) for simulations of tropical convection. They found in their different simulations total surface precipitation was affected by ice micro-physics parameterizations varying by about 13 percent.

Later many studies showed cumulus schemes to be critical and sensitive even in high resolution cloud resolving models (Brown and Swann, 1997; Wang, 2002). As computer capacity was not good enough to simulate numerical models in high resolutions, only warm-rain cloud microphysics was considered in the early tropical cyclones simulations (Jones, 1980; Rosenthal, 1978; Yamasaki, 1977). Although with simulation running on very high resolutions it is as yet unclear whether and to what degree the simulated tropical cyclone structure, intensification, and intensity can be affected by using different cloud microphysics parameterization schemes (Nasrollahi et al., 2012; Wang, 2002).

As a Case study for checking the impacts and sensitivity of different cumulus physics tropical Cyclone Viyaru was taken. Viyaru was a rapid intensifying tropical cyclone that made landfall in Chittagong, Bangladesh on the 16th of May, 2013. It was category 01B storm which formed May 10, 2013. Origin of cyclone Viyaru was a low pressure area in the Bay of Bengal which threatened Myanmar and Bangladesh with

fahadabdullaah@gmail.com

the speed of the wind about 100 kph. Cyclonic Storm Viyaru was located near latitude 21 degree N and longitude 90 degree E, about 240 km (150 mi) south-southeast of Kolkata and 240 km southwest of Chittagong in 16 May 2013 (Kotal and Bhattacharya, 2013).

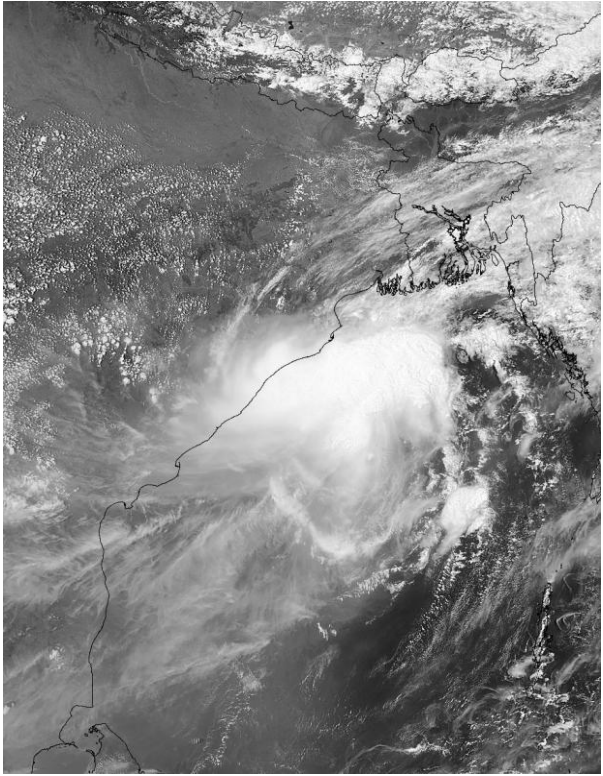

Figure 1: tropical cyclone Viyaru over Bay of Bengal, Image simulated by NASA

It was moving north-east wards. By evening of 16 May it was expected to cross the Bangladesh coast close to Chittagong. Maximum surface wind speed was estimated at 85 km/h (53 mph). Here the storm had extensive cloud mass, which brought unsettled weather to Sri Lanka, Thailand, and south eastern India. Even before the storm hit at least 18 deaths relating to Viyaru had been reported in Bangladesh, Burma and Sri Lanka. After the storm outs beside crops about 7,500 houses were damaged in the cyclone. Right after Cyclone Viyaru hits Bangladesh, 45 people were killed. Viyaru was downgraded as it made landfall to a category one cyclone and was given a threat level of seven on a 10-point scale. The storm was relatively weaker than expected because of the landfall in Chittagong Bangladesh. It could damage more than it did and kill lots of people in process if it was not downgraded and get weaker.

## 2. Experiment Design and Methods

### 2.1 Model Set- up

Tropical cyclone Viyaru was simulated by WRF-ARW v3.4.1 in a nested domain. The model domain consists of one parent domain and one nested domain. The resolution of the parent domain is 36 km while the nested domain has resolution of 12 km. The central point of the domain's latitude is 17.59 and longitude 89.58, with NCEP Global Final Analysis (FNL) data as initial and boundary conditions with 1 degree data resolution.

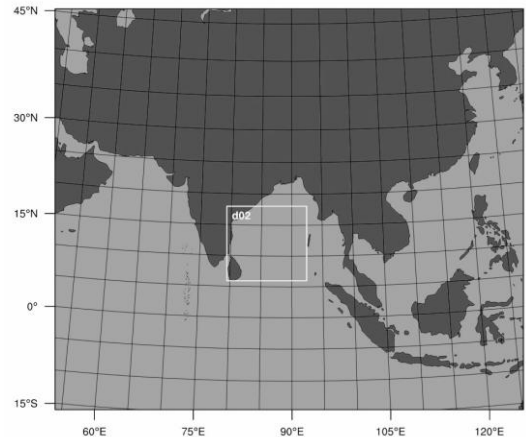

**Figure 2:** Parent and nested domain used for simulation

Type of data set used for this simulation is ds083.2 fnl, WMO GRIB1. Start date of the data set was 2013-05-11 18:00 and end date of the data set 2013-05-16 18:00. Geog data resolution used for both parent and nested domain is 10m. Map projection of the model was lambert. Data sets vertical lever was 35 with no of grid points we=230,118, sn=200,112; for parent and nested domain respectively.

### 2.2 Physics Sensitivity Simulation

For the best result data of 11 May to 16 May of 2013 were simulated where the cyclone Viyaru were most intensified at 13 may to 14 may of 2013. Four numerical simulations was done with same microphysics scheme (WSM3), planetary boundary layer scheme (YSU), NOAH land Surface scheme but different Cumulus Parameterization scheme. Common configuration and model physics of the WRF used for all simulation is shown in table 1.

**Table1:** configuration used for WRF model in simulations

| Options | Used |
|---|---|
| WRF model | ARW-wrf 3.4.1 |
| Maximum domain | 2 |
| Domain resolution | 36 for parent and 12 for nested |
| Time step | 120 |
| Vertical levels | 35 |
| mp_physics | 3 (WSM3)for both domain |
| radt | 36 for parent and 12 for nested domain |
| pbl | 1 (YSU) |
| cudt | 5 mins |

fahadabdullaah@gmail.com

For four cases of numerical simulations different cumulus physics were used but were same for the both parent and nested domain at same time. Cumulus physics schemes fall into two main classes. Adjustment type cumulus schemes and mass flux type cumulus physics. Betts-Miller-Janjic or BMJ is adjustment type which relaxes towards a post convective sounding. All other cumulus schemes are mass flux type which determines updraft and sometimes downdrafts mass flux and other fluxes.

**Table2:** simulation numbers with different CU scheme options

| Simulation number | Cu phy | Scheme | Momentum Tendencies | Shallow Convection |
|---|---|---|---|---|
| (a) | 0 | Without any scheme | no | no |
| (b) | 1 | Kain-Fritsch (KF) | no | yes |
| (c) | 2 | Betts-Miller-Janjic(BMJ) | no | yes |
| (d) | 3 | Grell-Freitas | no | yes |
| (e) | 6 | Modified tiedtke | no | Yes |

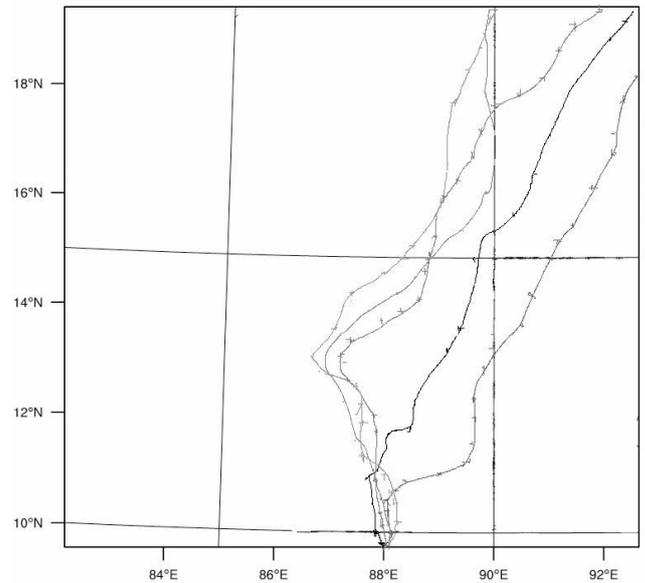

**Figure 3:** simulated track from WRF of domain 2 output

For all simulation different combinations were used is shown in table 2. First simulation (a) is without any cumulus physics option and other simulations (b), (c), (d) and (e) is with cumulus physics scheme Kain-Fritsch, Betts-Miller-Janjic, Grell-Freitas, and Modified Tiedtke respectively.

As the sea level pressure output was not significant enough to compare with observed data in case of simulation (e), modified Tiedtke simulation was used only for tracking purpose. The resulting output from the different physics combination simulations were analyzed and contrasted by calculating:

1. Time of landfall from the observed may 16, 2013.
2. Location of landfall from the actual location at Bay of Bengal (in km).
3. The deviation in landfall central pressure from the observed data (in hPa).
4. To calculate perfect simulation configuration these values were used to make an error index for each simulation. In terms of landfall pressure, timing and location The lowest resultant error index would indicate the most accurate simulation.

To get most accurate simulation compared to observation, data provided by Joint Typhoon Warning Centre (JTWC) and NASA tropical cyclone data was used. Figure 4 shows the observed tracking path of the cyclone with respect to different time and date.

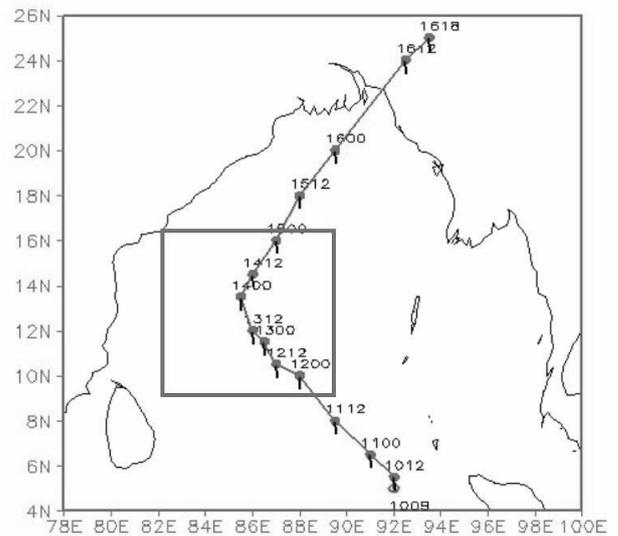

**Figure4:** Observed track of cyclone Viyaru.

**Tracking error with observed data**

After comparing with observed data, tracking error found by different simulations were counted and Normalized root mean square (RMS) error of each simulation was taken.

## 3. Results

**Tracks**

Figure 3 shows the output tracks from different simulations. Although at primary stage all simulation seems to be clustered around same place but with time different cumulus physics gave different tracking results with different time period. From the figure simulation number (e) with cu 6 seems to gives most accurate result with track respect with time period.

fahadabdullaah@gmail.com

**Table 3:** RMS error of 5 simulations

| Simulation number | Normalised root-mean-square (RMS) error of each simulation |
|---|---|
| (a) | 0.97 |
| (b) | 0.61 |
| (c) | 0.87 |
| (d) | 0.68 |
| (e) | 0.49 |

From graph (figure 5) and table 3 we can see that simulation (e) has least error while simulation (a) gives the most distracted path of the cyclone Viyaru.
Modified Tiedke scheme gives better track positions, though kain-Fritch scheme also generates good outcomes with less vector displacement and landfall errors. Although cumulus scheme Grell-Freitas is also mass flux type like KF and Modified Tiedke, it has drawbacks as it is not designed for elevated convection.

On the other hand simulation (c) with cumulus scheme Betts-Miller-Janjic (BMJ) gave track error near as simulation (a) with no cumulus physics schemes.

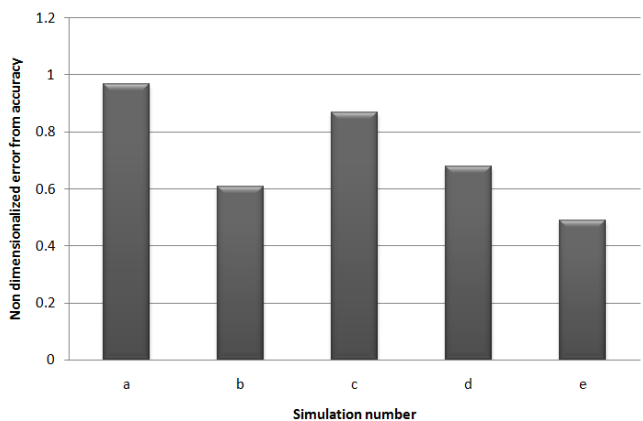

**Figure 5:** Graph showing (RMS) error of each simulation.

**Precipitation and sea level pressure**

However taking tracking and landfall location as major factor to determine a simulation as most accurate is not good idea. If we compare Total precipitation and sea level pressure with different run of simulation, CU 1 (simulation number (b)) gave most accurate result with observed data. All simulation from the WRF model had same physics configuration except cumulus physics scheme. In simulation number (b) cyclone Viyaru has fairly realistic value compared to observed data.
On primary stage, cyclone Viyaru pressure level increases on Open Ocean but after landfall with the pressure starts decreasing. Compared to simulation (b), other simulations result with sea level pressure tends to give more error with time.

Figure 6 shows total precipitation (mm) on Bay of Bengal from different simulations, which is initiated on 11 may 2013 through 14 may 2013 for all four runs.

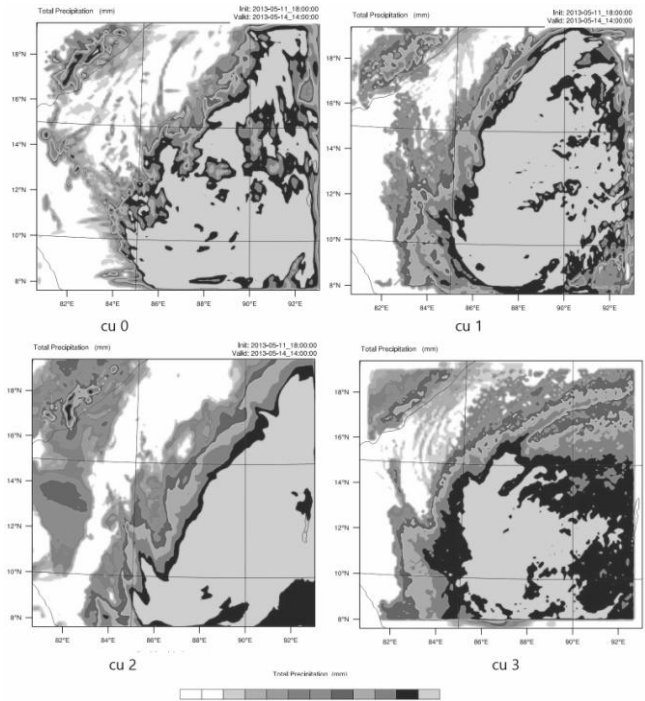

**Figure 6:** Total precipitation (mm) on Bay of Bengal from different simulations

Figure 7 shows a simulated 3-D analysis of NASA´s Tropical Rainfall Measuring Mission (TRMM) satellite´s multi satellite Precipitation Analysis (TMPA).

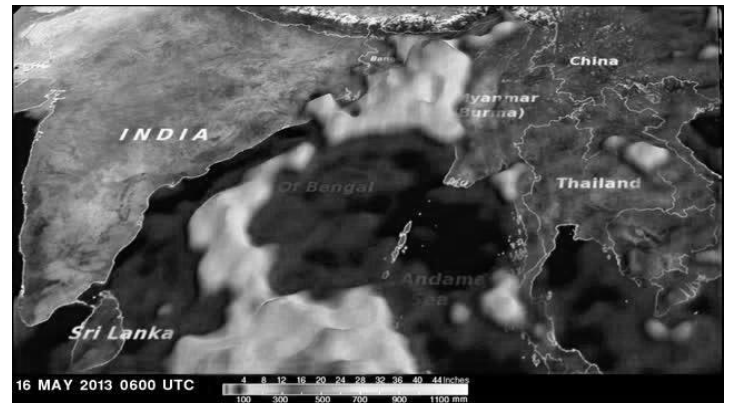

**Figure 7:** Total precipitation from NASA 3-D simulated analysis

It shows total precipitation during the time tropical cyclone Viyaru was making deadliest transit through Bay of Bengal with initiated date 6 may 2013 to end date 16 may 2013. Over Bangladesh and north-east India, TRMM's Precipitation Radar (PR) found rain within Viyaru falling at a rate of over 67mm/hr (~2.6 inches) on May 15 and at a rate of over 57mm/hr (~2.25 inches) on May 16. With this analysis total precipitation of about 544mm (~21.4 inches) were found.

After studying the precipitation tendency (figure 8) and sea level pressure it seems that without any shallow convection update from cumulus physics simulation (a) (without cumulus physics scheme) tends to give least significant result. Simulation (b) tends to give more finest and realistic result compared with observed data for intensity and central pressure

fahadabdullaah@gmail.com

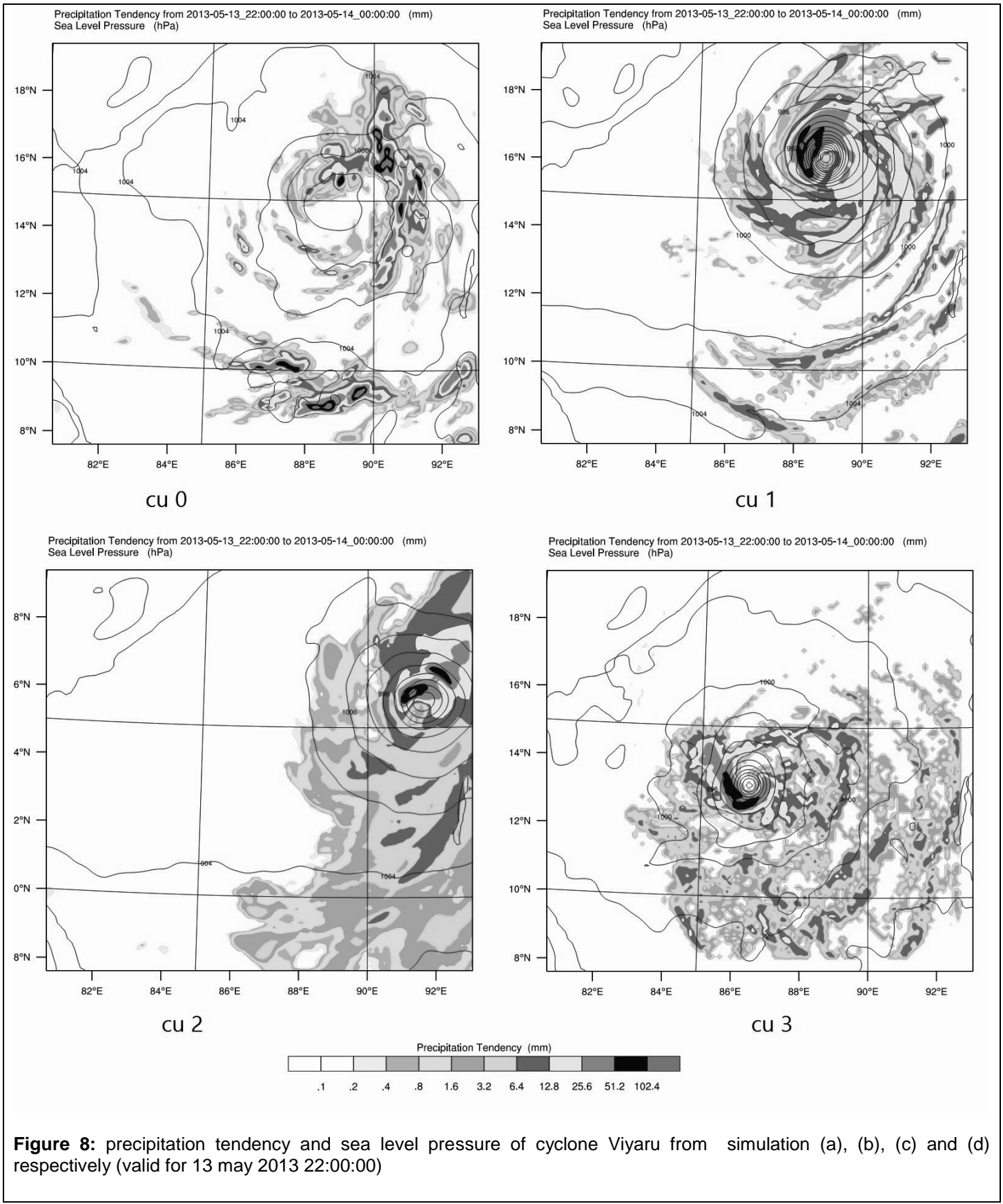

**Figure 8:** precipitation tendency and sea level pressure of cyclone Viyaru from simulation (a), (b), (c) and (d) respectively (valid for 13 may 2013 22:00:00)

fahadabdullaah@gmail.com

with explicit precipitation than the other simulations. The reason behind simulation (b) works fine with central pressure and precipitation may occur for triggering of convection sharply by Kain-Fritsch scheme.

On the other hand simulation (a) and (c) doesn't produce good result with precipitation. As simulation (a) is without any cumulus physics scheme and in case of simulation (c) Betts-Miller-Janjic (BMJ) was used. As BMJ scheme has issue with moisture in soundings being not enough, was the reason behind giving lacks in output compared to observed data.

Figure 8 shows precipitation tendency (mm) and sea level pressure (hPa) on 13 may 2013 at 22:00:00 to 14 may 2013 at 00:00:00 for different simulations. The precipitation tendency and sea level pressure was taken from the same time 13 may 2013, 22:00:00.

In terms of lowest central pressure simulation (b) and (d) was accurate enough to produce nearly realistic result.

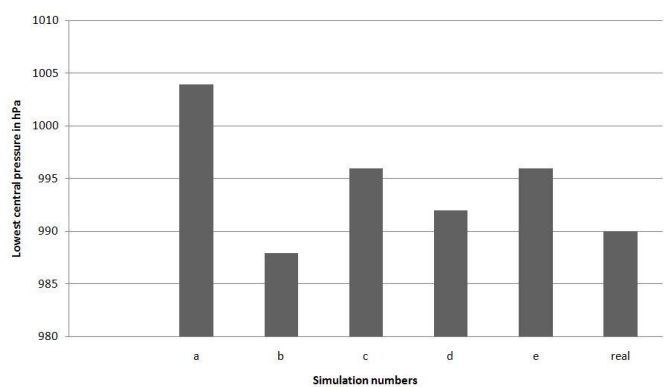

**Figure 9:** Simulated lowest central pressure of cyclone over Bay of Bengal compared with real observed data

However simulation (a) gave least accurate value of central pressure of cyclone viyaru, through 10 may to 16 may 2013.

## Conclusion:

Studying the simulated data conclusion can be easily made that, taking notice of shallow convection with cumulus physics parameterization was better than without. As a result without any cumulus physics scheme, simulation gave the least accurate result with sea level pressure and total precipitation with ambiguous results regarding track of cyclone. On the other hand mass flux type (KF, GF) cumulus scheme had better result than adjustment type cumulus scheme (BMJ), as BMJ has key issue with lacking of moisture in soundings.

Conclusion can be drawn from the results that simulating a cyclone or typhoon event accurately is vastly dependent on many parameters such as domain set up, map projection, boundary conditions and different results with different packages of physics for the model.

With cumulus physics scheme Kain-Fritsch the most accurate pressure value of sea level and changes with respect of time was found and cumulus physics scheme Modified Tiedke gives the most accurate track with least landfall error. If we are looking for a comprehensive result from a cumulus scheme Kain-Fritsch is our answer. Although for best and most accurate result further studies will be needed with sensitivity of other physics packages such as planetary boundary level (pbl) and microphysics scheme.